\documentclass{article}

\usepackage{amsmath,latexsym,amssymb}

\usepackage{theorem}
\newtheorem{definition}{Definition}[section]
\newtheorem{lemma}[definition]{Lemma}
 
\newtheorem{theorem}[definition]{Theorem}

\def\qed{\ifmmode$\Box$\else{\unskip\nobreak\hfil
\penalty50\hskip1em\null\nobreak\hfil$\Box$
\parfillskip=0pt\finalhyphendemerits=0\endgraf}\fi}
\def\endenv{\ifmmode\;\else{\unskip\nobreak\hfil
\penalty50\hskip1em\null\nobreak\hfil\;
\parfillskip=0pt\finalhyphendemerits=0\endgraf}\fi}
\newenvironment{proof}{\noindent \textbf{{Proof~} }}{\qed}

\mathchardef\ordinarycolon\mathcode`\:     
\mathcode`\:=\string"8000
\def\vcentcolon{\mathrel{\mathop\ordinarycolon}}
\begingroup \catcode`\:=\active
  \lowercase{\endgroup
  \let :\vcentcolon
  }

\newcommand{\nc}{\newcommand}
\nc{\rnc}{\renewcommand}
\nc{\beq}{\begin{equation}}
\nc{\eeq}{{\end{equation}}}
\nc{\beqa}{\begin{eqnarray}}
\nc{\eeqa}{\end{eqnarray}}
\nc{\B}{\cal{B}}
\nc{\lbar}[1]{\overline{#1}}
\nc{\bra}[1]{\langle#1|}
\nc{\ket}[1]{|#1\rangle}
\nc{\ketbra}[2]{|#1\rangle\!\langle#2|}
\nc{\braket}[2]{\langle#1|#2\rangle}
\nc{\proj}[1]{| #1\rangle\!\langle #1 |}
\nc{\inner}[2]{(#1,#2)}
\nc{\IP}{\operatorname{IP}}
\rnc{\max}{\operatorname{max}}
\nc{\Prob}{\operatorname{Prob}}
\nc{\rank}{\operatorname{rank}}
\nc{\Span}{\operatorname{Span}}
\rnc{\S}{\operatorname{S}}
\nc{\diag}{\operatorname{diag}}
\nc{\sign}[1]{\mbox{sign$(#1)$}}
\nc{\smfrac}[2]{\mbox{$\frac{#1}{#2}$}}
\nc{\Spec}{\operatorname{Spec}}
\nc{\Tr}{\operatorname{Tr}}
\nc{\xor}{\oplus}
\nc{\ox}{\otimes}
\nc{\dg}{\dagger}
\nc{\dn}{\downarrow}
\nc{\cA}{{\cal A}}
\nc{\cB}{{\cal B}}
\nc{\cC}{{\cal C}}
\nc{\cD}{{\cal D}}
\nc{\cE}{{\cal E}}
\nc{\cF}{{\cal F}}
\nc{\cG}{{\cal G}}
\nc{\cH}{{\cal H}}
\nc{\cI}{{\cal I}}
\nc{\cJ}{{\cal J}}
\nc{\cK}{{\cal K}}
\nc{\cL}{{\cal L}}
\nc{\cR}{{\cal R}}
\nc{\cS}{{\cal S}}
\nc{\cX}{{\cal X}}
\nc{\supp}{{\operatorname{supp}}}
\nc{\stab}{{\operatorname{stab}}}
\nc{\rar}{\rightarrow}
\nc{\lrar}{\longrightarrow}

\def\a{\alpha}
\def\b{\beta}

\def\e{\epsilon}

\def\l{\lambda}

\def\r{\rho}
\def\s{\sigma}

\def\ph{\varphi}

\def\ps{\psi}
\def\om{\omega}

\def\L{\Lambda}

\def\S{\Sigma}

\def\Ph{\Phi}

\nc{\RR}{{{\mathbb R}}}
\nc{\CC}{{{\mathbb C}}}
\nc{\FF}{{{\mathbb F}}}
\nc{\NN}{{{\mathbb N}}}
\nc{\ZZ}{{{\mathbb Z}}}
\nc{\PP}{{{\mathbb P}}}
\nc{\QQ}{{{\mathbb Q}}}
\nc{\UU}{{{\mathbb U}}}
\nc{\Q}{{{\bar{Q}}}}

\title{\textbf{R\'enyi-entropic bounds on quantum communication}}
\author{Wim van Dam\thanks{Computer Science Division, Soda Hall, 
University of California, Berkeley, CA 94720 (USA). 
Also at MSRI Berkeley and HP Palo Alto.
Email: vandam@cs.berkeley.edu.}
\and Patrick Hayden\thanks{Institute for Quantum Information, Caltech 107-81, 
Pasadena, CA 91125 (USA). Email: patrick@cs.caltech.edu.}}

\begin{document}
\maketitle
\begin{abstract}
In this article we establish new bounds on the quantum 
communication complexity of distributed problems.
Specifically, we consider the amount of communication 
that is required to transform a bipartite state 
$\ph_{AB}$ into another, typically more entangled, $\ps_{AB}$.
We obtain lower bounds in this setting 
by studying the \emph{R\'enyi entropy} of the marginal 
density matrices of the distributed system.

The communication bounds on quantum state transformations 
also imply lower bounds for the model of communication
complexity where the task consists of the the distributed evaluation of 
a function $f(x,y)$.
Our approach encapsulates several known lower bound 
methods that use the log-rank or the von Neumann entropy
of the density matrices involved.  The technique is also effective
for proving lower bounds on problems involving a promise or
for which the ``hard'' distributions of inputs are correlated.
As examples, we show how to prove
a nearly tight bound on the bounded-error
quantum communication complexity of the inner product function
in the presence of unlimited amounts of EPR-type entanglement
and a similarly strong bound on the complexity of the shifted
quadratic character problem. 
\end{abstract}

\section{Introduction}
Quantum information theory investigates the power and limitations
of quantum mechanics for communicating and processing information.
In this article we look at the usefulness of quantum
communication for the execution of distributed tasks
in the presence of prior entanglement.
More specifically, we will focus on the question how many 
quantum bits (qubits) two parties $A$ and $B$ need to exchange 
in order to create a joint quantum state $\ps_{AB}$.
Our main tool will be the R\'enyi entropy of the marginal 
density matrix $\ps_A := \Tr_B(\ps_{AB})$.  

\subsection{Relation with Earlier Work} 
The quantum communication complexity of distributed state 
preparation has been discussed in \cite{ClevevNT99}.
In this article the authors gave a tight characterization
of the number of bits that two parties need to exchange in
order to approximate a joint state $\ps_{AB}$.
This characterization relied critically on the assumption
that initially $A$ and $B$ do not share any entanglement.

The setting in which the parties are allowed to 
use entanglement to assist their state preparation 
has been the topic of many investigations in the physics
community.  Typically, these articles concentrated on
the possibility of creating $\ps_{AB}$ 
under the restriction that only classical communication
is allowed~\cite{BBPS:cpeblo,JonathanP99,VidalJN00}.

The quantum communication complexity of distributed functions
was first considered in the work by Yao and 
Kremer~\cite{Y:qcc,K:qc}.  Kremer's work in particular
demonstrated several lower bounds on the quantum 
communication complexity of distributed functions. 
The first example of a quantum
protocol that requires less communication than any 
classical solution was described by Cleve and Buhrman 
in 1997~\cite{CB:sqefc}. 
General lower bound techniques for quantum communication have been
described have been described in 
\cite{Ambainis98}, \cite{BW}, \cite{ClevevNT99}, 
\cite{Klauck01}, \cite{NayakS} and \cite{Razborov02}.  
Specifically, the results in this paper can be interpreted as
generalizing and refining the lower bound techniques of
\cite{Ambainis98} and \cite{ClevevNT99}.

\subsection{Tools from Quantum Information}
In this paper we will often be concerned with pure states 
$\psi_{AB} = \proj{\psi}_{AB}$ of a composite system $AB$.
For any such state there exist orthonormal bases $\{\ket{i_A}\}$ on $A$
and $\{\ket{i_B}\}$ on $B$ such that 
$\ket{\psi_{AB}} = \sum_i \sqrt{\lambda_i} \ket{i_A} \ket{i_B}$.
This representation is known as the Schmidt decomposition.
The degree of entanglement of $\psi_{AB}$ is then naturally 
characterized by the spectrum $\{\lambda_i\}$ of the reduced density 
operator $\psi_A := \Tr_B(\psi_{AB})$.
(Since $\psi_{AB}$ is pure, $\psi_A$ and $\psi_B$ have the same spectra.)
In particular,  
$\log\operatorname{rank}(\psi_A)$ and the von Neumann entropy
$S(\psi_A) := - \Tr( \psi_A \log \psi_A )$
have been extensively studied as \mbox{``measures of entanglement''.}
Indeed, in the limit of large numbers of copies of the state $\psi_{AB}$,
$S(\psi_A)$ completely characterizes the states reversibly convertible
into $\psi_{AB}$ using only local operations and classical
communication \cite{BBPS:cpeblo}.

Since we will be be studying approximate transformations of quantum
states and their connection to bounded-error quantum communication
complexity, it will be helpful to introduce some distance measures
from quantum information theory.
The \emph{Uhlmann fidelity} $F$ between two states $\rho$ and 
$\sigma$ is defined by
$F(\rho,\sigma) =  
\Tr ( \r^{1/2} \s \r^{1/2} )^{1/2}$.
The range of the Uhlmann fidelity is the interval $[0,1]$, reaching
the extreme values $0$ and $1$ for perfectly distinguishable and
identical states, respectively.
In the case of two pure states $\phi$ and $\psi$, the
expression for $F$ simplifies to $F(\phi,\psi) = |\braket{\phi}{\psi}|$.
Thus, the fidelity $F(\rho,\sigma)$ indicates how `close' 
$\rho$ and $\sigma$ are.  Furthermore, no quantum mechanical 
evolution decreases the fidelity between two states: 
$F(\rho,\sigma) \geq F(\L(\rho),\L(\sigma))$
for all states $\rho,\sigma$ and all trace-preserving, 
completely positive mappings $\L$ \cite{U:tp,J:ffmqs}.
In particular, if $\L$ is the partial trace $\Tr_B$, this reduces
to 
$F(\rho_A,\sigma_A)\leq F(\rho_{AB},\sigma_{AB})$
for any bipartite states $\rho_{AB}$ and $\sigma_{AB}$. 


\section{Quantum Communication Complexity of 
State Transformation and Distributed Computation}

Our starting point will be the question of how much 
communication between two parties $A$ and $B$ 
is required to transform a joint quantum state 
$\ph_{AB}$ into another state $\ps_{AB}$.
To distinguish between the settings where $A$ and $B$
are or are not allowed to use prior entanglement, we
introduce the following three complexity measures:
 
\begin{definition}
The \emph{transformation complexity},
$\Q_\e({\ps_{AB}}|{\ph_{AB}})$, is the minimum number
of bits (quantum and classical) of communication required, starting 
from the known state
$\ket{\ph_{AB}}$ to produce an approximating state ${\ps}'_{AB}$ such that
$F(\ps_{AB},{\ps}'_{AB}) > 1-\e$. 
\end{definition}
When $\e=0$ this definition requires that the final state $\ps'_{AB}$ be
equal to $\ps_{AB}$.  It is also worth emphasizing that 
this definition of $\Q_\e$ does not allow $A$ and $B$ to 
have any prior entanglement other than that of their state $\ph$.
The following complexity definition does allow such a resource:
\begin{definition}
The \emph{entanglement-enhanced transformation complexity} 
$\Q_\e^*$ of the two states $\ps_{AB}$ and $\ph_{AB}$
is defined by
\begin{eqnarray}
\Q_\e^*(\ps_{AB}|\ph_{AB})
&=& \inf_{{\om_{AB}}} \Q_\e(\ps_{AB} \ox \om_{AB}|\ph_{AB} \ox \om_{AB}),
\end{eqnarray}
where the infimum is taken over all possible bipartite pure states
$\ket{\om_{AB}}$. 
\end{definition}
A third version of transformation complexity concerns the usage
of prior entanglement only in the form of `EPR-pairs'.
\begin{definition}
The \emph{EPR-enhanced transformation complexity} $\Q_\e^{(*)}$ 
is defined by 
\begin{eqnarray}
\Q_\e^{(*)}(\ps_{AB}|\ph_{AB})
&=& \inf_{{\Phi^k_{AB}}} \Q_\e(\ps_{AB} \ox \Phi^k_{AB}|\ph_{AB} \ox \Phi^k_{AB}),
\end{eqnarray}
where the infimum is taken over all maximally entangled states of any rank \\
$\ket{\Phi^k_{AB}} := \smfrac{1}{\sqrt{k}}\sum_{j=1}^k\ket{j,j}$.
\end{definition}
An important feature of the two entanglement-enhanced complexities
is the requirement that the entangled state $\om_{AB}$ (or $\Phi^k_{AB}$)
be returned (with fidelity at least $1-\e$) 
after the transformation protocol.
Without this requirement the definitions would be vacuous because in
the setting where we are allowed to use arbitrary entangled states
as a disposable resource we could always choose $\om=\ps$ such that
no communication is necessary.

Instead, we view the prior entanglement as a potential 
\emph{catalyst}, which allows us to reduce the communication complexity
of the state transformation without destroying the entanglement.
The fact that such catalysis is possible in a non-trivial way was first
pointed out in \cite{JonathanP99}.

We will also make use of some shorthand notation to describe
state transformations in which $A$ and $B$ do
not share any initial state $\ph$. 
The complexities of such \emph{state preparations} will be denoted 
without the conditional $\ph$: $\Q_\e(\ps), \Q^*_\e(\ps)$ and 
$\Q^{(*)}_\e(\ps)$.


The `quantum sampling complexity' $\stackrel{\bullet}{Q}_\e(\ps)$ that was discussed 
in \cite{Ambainis98} concerned the (approximate) preparation of quantum 
states without the use of prior entanglement. 
Hence the $\stackrel{\bullet}{Q}_\e(\ps)$ complexity in that article 
equals the quantity $\Q_\e(\ps)$ that we defined here.

In the next sections of this article we will prove a dramatic 
difference between the complexities $\Q^*_\e$ and $\Q^{(*)}_\e$.
On the one hand, we will show that $\Q^*_\e$ is always 
zero for every $\e>0$, while on the other hand we will 
describe several lower bound techniques for the
complexity $\Q_\e^{(*)}$. 

The lower bounds on the EPR-enhanced transformation complexity 
will also give rise to several lower bounds on the
\emph{quantum communication complexity} for the distributed
evaluation of Boolean functions $f:X\times Y \rightarrow \{0,1\}$.
In this setting, Alice receives an input $x \in X$ and Bob 
gets $y \in Y$.  The question then is how many bits
 Alice and Bob need to communication
for a (probabilistic) evaluation of the function value $f(x,y)$.
The original version, in which only classical communication is
allowed, was first introduced by Yao \cite{Yao79} and has
developed rapidly since \cite{KushilevitzN97}.  In the quantum
context, it makes sense to allow Alice and Bob various
combinations of quantum resources to complete the task, such as
communication of qubits or pre-existing entanglement. More
specifically, let us assume that Alice and Bob receive $n$ bit
inputs, so that $X=Y=\{0,1\}^n$ and that there is a family of
functions $f$, one for each $n$. We write $Q_\e^*(f)$ to denote
the number of qubits of communication, as a function of $n$,
required to evaluate $f$ with probability of success at least
$1-\e$ on every input if Alice and Bob are allowed arbitrary
pre-existing entanglement. We also write $Q_\e^{(*)}(f)$ for the
same function if Alice and Bob are initially allowed to share any
maximally entangled state but not other types of entanglement.

The connection between the communication complexity of 
distributed functions and that of state transformations
was first discussed in \cite{ClevevNT99}, where it was used to obtain 
a lower bound on the quantum communication complexity
of the inner product function.
We reproduce from \cite{Ambainis98} 
(with a small correction) the following version of the lower bound.
In the context of pre-existing entanglement, the connection is
between distributed function evaluation protocols that \emph{consume}
entanglement and state transformation protocols that are \emph{catalyzed}
by entanglement.

\begin{lemma}[Transition lemma]
\label{Elem:transition}
Let $f:X\times Y \rightarrow \{-1,+1\}$ be a (partial) function
with quantum communication complexity $Q^*_\e(f)$. For
any $\ell_2$ distribution $\a:X\times Y\rightarrow \CC$ that has $\a_{xy}=0$
where $f(x,y)$ is not properly defined and $\sum_{xy} |\a_{xy}|^2=1$,
we define the two states $\ket{\ph_{AB}} =
\sum_{(x,y)}{\alpha_{xy}\ket{x,y}}$ and $\ket{\ps_{AB}} =
\sum_{(x,y)}{\alpha_{xy}f(x,y)\ket{x,y}}$. The transformation
complexity of these states is bounded from above by
$\Q^*_{2\e}(\ps|\ph)\leq 2Q^*_\e(f)$.  Similarly,
$\Q^{(*)}_{2\e}(\ps|\ph) \leq 2Q^{(*)}_\e(f)$ and
$\Q_{2\e}(\ps|\ph) \leq 2Q_\e(f)$.
\end{lemma}
\begin{proof}
Suppose that we have an $m$-bit protocol to compute $f$ with error
probability at most $\e$.  
First, $A$ and $B$ run this protocol on the distributed distribution
$\ph_{AB}$, which will produce the superposition
$\sum_{(x,y)}{\alpha_{xy}\ket{x,y,f(x,y),g_{xy}}}$ where 
$g_xy$ is the garbage that the protocol has produced. 
Next, the function value $f(x,y)$ is used for the phase changing
operation $\ket{x,y} \mapsto f(x,g)\ket{x,y}$.
Finally, the $m$-bit protocol is executed `in reverse' to 
erase the garbage qubits $\ket{f(x,y),g_{xy}}$. 
The final result of this $2m$-bit protocol
 is with fidelity $1-2\e$ the desired state 
$\ket{\ps}  = \sum_{(x,y)}{\alpha_{xy}\ket{x,y}}$.

See the appendix of \cite{Ambainis98} for a more detailed analysis of the 
fidelity of this procedure. 
(Note however that fidelity error of the resulting $\Q$-protocol
is $2\e$, not $\e$ as mentioned in 
\cite{Ambainis98}[Ambainis, private communication].)
\end{proof}

\section{Some Basic Observations}
\subsection{Embezzling entanglement}
While the distinction drawn in the previous section between the
functions $\bar{Q}_\e^*$ and $\bar{Q}_\e^{(*)}$ might appear
to be academic at first, their behaviors are actually radically different:
$\bar{Q}_\e^*$ is identically zero for $\e\neq 0$ but $\bar{Q}_\e^{(*)}$
is not.  In other words, transformation complexity is trivial in the
presence of arbitrary entanglement but not in the presence of
unlimited amounts of maximal entanglement.
The reason for this striking difference lies in the remarkable properties
of the family of states
\begin{eqnarray}
\ket{M(d)} &:=& \frac{1}{\sqrt{H_d}} \sum_{j=1}^d
\frac{1}{\sqrt{j}} \ket{j_A} \ket{j_B},
\end{eqnarray}
where $H_d := \sum_{j=1}^d \frac{1}{j}$ is chosen so that
$\ket{M(d)}$ is normalized.  
\begin{theorem}[Embezzlement \cite{embezzle}] \label{thm:embezzle}
If $\e>0$ is fixed and $\ket{\ph_{AB}}$ is 
any entangled state, then for all sufficiently large $d$,
$\bar{Q}_\e(M(d)\ox \ph_{AB}|M(d)) = 0$.  
Consequently, $\bar{Q}_\e^*$ is zero for all $\e>0$.
\end{theorem}
Thus, it is possible to \emph{embezzle} a copy of $\ket{\ph_{AB}}$
from $\ket{M(d)}$, thereby removing a small amount of entanglement from the
original state, while causing only an arbitrarily small disturbance $\e$ 
to it.  See \cite{embezzle} for a proof of this theorem.

\subsection{One-round versus Multi-round Protocols}
\begin{lemma}
\label{lem:oneRound}
If there exists a bipartite quantum communication protocol that 
transforms the initial state $\ket{\ph_{AB}}$ to final state 
$\ket{\psi_{AB}}$ that requires $q$ qubits of communication, then there
also exists a one-round protocol, $q$ qubit protocol that
establishes the maping $\ket{\ph_{AB}}\mapsto \ket{\ps_{AB}}$.
Hence, the transformation complexities $\bar{Q}$ and 
$\bar{Q}^{(*)}$
are completely 
described by the complexities of the one-round protocols. 
\end{lemma}
\begin{proof}
First we note that at any stage of the protocol, 
$A$ and $B$ always have perfect knowledge about the state 
that they share at that moment.  This is due to the 
facts that $A$ and $B$ know the initial state $\ph_{AB}$
and that the procotol can be assumed to be unitary. 
Specifically, it is always known what the Schmidt coefficients
are of the current state.  With this viewpoint it is clear that 
the communication only serves the purpose of changing these
coefficients, everything else can be done locally.
A qubit of communication from Bob to Alice will thus
establish a transformation 
$\sum_{i}{\sqrt{\lambda_i}\ket{i_A,i_B}} 
\mapsto \sum_{i}{\sqrt{\lambda'_i}\ket{i'_A,i'_B}}$,
written in the Schmidt decomposition of the prior and posterior
state.  By symmetry, this same transformation 
$\lambda_i \mapsto \lambda'_i$ can also be implemented by a
qubit of communication from Alice to Bob.
Hence, in general, all communication from Bob to Alice can be
replaced by an equal amount of communication in the opposite 
direction.
\end{proof}

This argument is very similar to the one given by Lo and 
Popescu~\cite{LoP01} to reduce the study of multi-round LOCC
(Local Operations and Classical Communication) state
transformation protocols to single round protocols.

This lemma also points to a connection with another basic
question about quantum states.  Suppose Alice and Bob initially share
a tripartite state $\ket{\ph_{ACB}}$, with Alice in possession of
$A$ and $C$. If Alice sends subsystem $C$ to Bob, then her reduced
density operator changes from $\ph_{AC}$ to $\ph_A$. Thus, the
problem of state transformation is intimately connected to the
question of determining the effect of a partial trace operation
on a density operator's spectrum.  While the general answer to
this question is unknown, Nielsen and Kempe \cite{NielsenK00}
have found some partial results for the case where $\ph_{AC}$ is separable.


\subsection{Difference between $\bar{Q}$ and $\bar{Q}^{(*)}$?}
Is there a difference between the transformation 
complexities $\bar{Q}$ and $\bar{Q}^{(*)}$?  Or, in other
words, does prior EPR entanglement help? 
We do not know the answer in the general case, but we do
get an indication that the two might be equivalent by looking
at the case where the prior state $\ph_{AB}$ is not 
entangled.  Recall that the transformation complexity
with a prior product state $\ket{\ph_{AB}} = 
\ket{\ph_A}\otimes\ket{\ph_B}$ is denoted by 
$\bar{Q}(\ps_{AB}) := \bar{Q}(\ps_{AB}|\ph_{A}\otimes\ph_B)$.
\begin{lemma}\label{lm:(*)}
$\bar{Q}_\e(\ps_{AB}) = \bar{Q}_\e^{(*)}(\ps_{AB})$. 
Hence, by the `quantum sampling' paper \cite{Ambainis98}, 
$\bar{Q}_\e^{(*)}(\ps_{AB}) = \bar{Q}_\e(\ps_{AB}) =
\stackrel{\bullet}{Q}_\e(\ps_{AB}) = \log(\rank_{1-\e}(\ps_A))$,
where $\rank_{1-\e}(\ps_A)$ is the size of the smallest subset of
eigenvalues of $\ps_A$ summing to at least $1-\e$.
\end{lemma}  
\begin{proof}
Let $(\lambda_1,\dots,\lambda_r)$ be the spectrum of the Schmidt 
decomposition of $\ps_{AB}$.  Assume that the optimal EPR-assisited
protocol uses the $k$-level state 
$\ket{\Phi^k_{AB}} := \smfrac{1}{\sqrt{k}}\sum_{j=1}^k\ket{j,j}$,
such that the spectrum change of this protocol is described by
$(\smfrac{1}{k},\dots,\smfrac{1}{k}) \mapsto 
(\lambda_1,\dots,\lambda_r)\otimes (\smfrac{1}{k},\dots,\smfrac{1}{k})
\equiv (\smfrac{\lambda_1}{k},\dots,
\smfrac{\lambda_1}{k},\smfrac{\lambda_2}{k},\dots,\smfrac{\lambda_r}{k})$, 
where we reshufled the amplitudes in nonincreasing order.
To get an approximating spectrum with no more than $\e$ error to
this last state, we need a final state with rank at 
least $k\cdot\rank_{(1-\e)}(\ps_{AB})$.   
Because we start with a rank $k$ state $\Ph^k_{AB}$ we see that we 
need at least $\log(\rank_{(1-\e)}(\ps_{AB}))$ qubits of communication,
which is also an upper bound on $\stackrel{\bullet}{Q}_\e(\ps_{AB}) := 
\Q_\e(\ps_{AB})$ (see \cite{Ambainis98}).
\end{proof}

\section{Lower bounds on ${\bar{Q}_\e^{(*)}}$}
In this section we will use the properties of the R\'enyi 
entropies~\cite{Renyi61} to prove bounds on the
transformation complexity in the presence of maximally entangled
states.  

\subsection{R\'enyi entropy}
\label{Isubsec:Renyi}

\begin{definition}
\label{Idefn:Renyi}
Given a density matrix $\rho$ with spectrum
$\lambda_1\geq\lambda_2\geq\dots\geq\lambda_r$, its \emph{R\'enyi
entropy} $S_\alpha(\r)$ is defined to be
\begin{eqnarray}
S_\alpha(\rho) & = & \frac{1}{1-\alpha}\log(\Tr \rho^\alpha) \quad
 = \quad \frac{1}{1-\alpha}\log\left(\sum_{i=1}^r{\lambda_i^\alpha}\right)
\end{eqnarray}
for every $0<\alpha<\infty$, $\alpha\neq 1$. Furthermore, for the
values  $\alpha=0,1,\infty$ we define
$S_0(\rho) =  \log(\rank(\rho))$,
$S_1(\rho) = -\Tr(\rho\log\rho)$ and
$S_\infty(\rho) = -\log(\lambda_{1})$.
Note that $\lim_{x\rightarrow \alpha}{S_x(\rho)}=S_\alpha(\rho)$.
\end{definition}

We have already encountered $S_1$, which is just the von Neumann
entropy $S$.  The other exceptional orders, $0$ and $\infty$, are,
likewise, quantities that appear frequently in quantum information
theory and communication complexity. The following theorem summarizes
the basic properties of the R\'enyi entropies.
\begin{theorem}
\label{Ithm:RenyiProperties}
Let $\r$ be a density operator with spectrum
$\lambda_1,\dots,\lambda_r$.  The following properties hold for
all $\a$:
\begin{tabbing}
\quad\=1. The inequalities $0 \leq S_\alpha(\rho) \leq \log \rank(\rho)$ hold. \\
\>2. For $\a\neq 0$, the entropy is maximized if and only if $\lambda_i = \smfrac{1}{r}$. \\
\>3. The entropy is minimized only by pure states. \\
\>4. (Additivity) For two density matrices:
 $S_\alpha(\rho_1\ox\rho_2)=S_\alpha(\rho_1)+S_\alpha(\rho_2)$. \\
\>5. If $\alpha \leq \beta$ then $S_\alpha(\rho)\geq S_\beta(\rho)$. \\
\>6. (Schur concavity) If $\rho \succ \sigma$ then
$S_\alpha(\rho)\leq S_\alpha(\sigma)$.
\end{tabbing} 
\end{theorem}
The notation $\rho \succ \sigma$ means that spectrum of the operator
$\rho$ majorizes the spectrum of $\sigma$.  That is, if $\sigma$ has
eigenvalues $\gamma_1 \geq \gamma_2 \geq \dots \geq \gamma_r$ then
$\sum_{j=1}^l \l_j \geq \sum_{j=1}^l \gamma_j$ for all $1 \leq l \leq r$.

It is less well-known that the R\'enyi entropies obey a weak form of
subadditivity, which will be crucial for proving bounds on communication.
While subadditivity of the form $S(\r_{AB}) \leq S(\r_A) + S(\r_B)$
fails for the R\'enyi entropies unless $\a$ is $0$ or $1$, replacing
one of the terms on the right hand side by the logarithm of the
rank of $\r_B$, that is, by $S_0(\r_B)$, gives an inequality that
holds for all $\a$.

\begin{lemma}[Weak subadditivity]
\label{Ilem:weakSubAdd}
Let $\rho_{AB}$ be a bipartite density matrix. For all $\a$, the
R\'enyi entropy of the states is bounded by
\begin{equation*}
S_\alpha(\rho_A)-S_0(\rho_B) \quad \leq \quad
S_\alpha(\rho_{AB}) \quad \leq \quad S_\alpha(\rho_A)+S_0(\rho_B).
\end{equation*}
\end{lemma}
\begin{proof}
We will first prove the upper bound on $S_\alpha(\rho_{AB})$ and
then show how the lower bound follows. Let $\{\ket{i_A}\}$ be the
eigenbasis of the reduced density operator $\rho_A$ and consider
the projection
\begin{eqnarray*}
P(\rho_{AB}) &=& \sum_i \proj{i_A} \otimes  \bra{i_A} \rho_{AB}
\ket{i_A}.
\end{eqnarray*}
Since $P$ is a doubly stochastic mapping and the R\'enyi entropies
are all Schur-concave, we have $S_\alpha(\rho_{AB}) \leq
S_\alpha(P(\rho_{AB}))$.
We begin by assuming that $\alpha\neq 0,1,\infty$; the exceptional
values will follow by continuity.
Also, we will use the notation $\lambda_i = \Tr( \bra{i_A} \rho_{AB}
\ket{i_A} )$ and $\rho_i = \bra{i_A} \rho_{AB} \ket{i_A} /
\lambda_i$, such that $P(\rho_{AB}) = \sum_{i}{\lambda_i
\proj{i}\ox\rho_i}$.  
Next, observe that
\begin{eqnarray}
\sign{1-\alpha}\cdot\Tr(\rho_i^\alpha)& \leq&
\sign{1-\alpha}\cdot(\rank(\rho_i))^{1-\alpha}
\end{eqnarray}
for all $\alpha$, which is the last ingredient required
to finish the proof of the upper bound:
\begin{eqnarray}
S_\alpha(P(\rho_{AB}))
&=& \mbox{$S_\alpha (\sum_i \lambda_i \proj{i_A} \otimes \rho_i )$} \\
&=& \mbox{$\frac{1}{1-\alpha}
        \log\left(\Tr \left[\left( \sum_i \lambda_i \proj{i_A} \otimes \rho_i \right)^\alpha\right] \right)$} \\
&= & \mbox{$\frac{1}{1-\alpha}
        \log\left(\sum_i \lambda_i^\alpha \Tr \rho_i^\alpha\right)$} \\
&\leq& \mbox{$\frac{1}{1-\alpha}
        \log\left(\sum_i \lambda_i^\alpha (\rank(\rho_i))^{1-\alpha}\right)$} \\
&\leq& \mbox{$S_\alpha(\rho_A) + \log \rank(\rho_B)$}.
\end{eqnarray}
The last line follows from the fact that $\Tr_B(P(\rho_{AB})) =
\rho_A$ so that the $\lambda_i$ are just the eigenvalues of
$\rho_A$, and the fact that $\rank(\rho_i)\leq \rank(\rho_B)$.

The lower bound is a straightforward consequence of the upper
bound. Given $\rho_{AB}$, let $\psi_{ABR}$ be a pure state such
that $\Tr_R \psi_{ABR} = \rho_{AB}$, and hence $\Spec(\rho_A) =
\Spec(\psi_{BR})$ and $\Spec(\rho_{AB}) = \Spec(\psi_R)$.  The
upper bound reads $S_\alpha(\psi_{BR}) \leq S_\alpha(\psi_R)+\log
\rank(\psi_B)$, which is equivalent to
\begin{eqnarray}
S_\alpha(\rho_A) - \log \rank( \rho_B ) &\leq&
S_\alpha(\rho_{AB}).
\end{eqnarray}
The $\alpha = \infty$ case is proven via the  equality
$\lim_{x\rightarrow\infty}{S_x(\rho)} = S_{\infty}(\rho)$.
\end{proof}

There is an alternative method of proving lemma~\ref{Ilem:weakSubAdd}
that builds on known results about state transformations.  
Let $\ket{\ph_{ACB}}$ be
a tripartite pure state, with Alice in possession of the shares $A$
and $C$.  One way for Alice to send system $C$ to Bob would be to
first send him half a maximally entangled state of rank $k := \rank(\ph_C)$
and then to teleport \cite{teleportation} system $C$ using that 
maximally entangled state.
Since teleportation is an LOCC protocol, Alice's final reduced density
operator $\ph_A$ must majorize the operator 
$\ph_{AC} \ox \diag(1/k,\ldots,1/k)$ \cite{Nielsen}.  
Since the R\'enyi entropies 
are  Schur concave, we can deduce the left-hand inequality of~lemma
\ref{Ilem:weakSubAdd}.


\subsection{R\'enyi-entropic bounds on communication}
\label{Esubsec:Renyi}

The following lemma is the starting point for proving bounds on the
approximate transformation complexity.  Namely, it 
bounds the minimum number of qubits that need to be exchanged in order
to succeed in an exact transformation.

\begin{lemma}
\label{Elem:exactRenyi}
Let $\ket{\ph_{AB}}$ and $\ket{\ps_{AB}}$ be bipartite pure states.
If there exists a protocol for converting $\ket{\ph_{AB}}$ to
$\ket{\ps_{AB}}$ using $n$ qubits of communication, then
$n \geq S_\a(\ps_A) - S_\a(\ph_A)$, for all $\a$.
\end{lemma}
\begin{proof}
The most general protocol consists of $k$ rounds of communication,
separated by arbitrary completely positive, trace-preserving operations
performed locally by Alice and Bob.  Since each of these local operations
can be implemented by adjoining a pure ancilla state, applying a
unitary and then discarding part of the system, we can assume
without loss of generality that the first step of the protocol consists 
of adjoining pure, separable ancilla, that each local operation by
Alice or Bob is unitary and that the discard steps are all postponed to
the end of the protocol.  We will now analyze each of these three
phases in turn.  Adjoining separable ancilla does not affect the
non-zero part of the spectrum of the reduced density operator, or,
therefore, the R\'enyi entropy.  The second phase, by lemma 
\ref{lem:oneRound}, can be assumed to consist of a single round of
communication from Alice to Bob.  By the weak subadditivity of the Renyi
entropy, the increase in $S_\a$ at this step is no more than the
number of qubits of communication.  Finally, at the discard step,
by the additivity of the R\'enyi entropies over tensor products,
$S_\a$ is non-increasing.
\end{proof}

While lemma~\ref{Elem:exactRenyi} is itself a consequence of the
Schur concavity of the R\'enyi entropies and Nielsen's 
theorem~\cite{Nielsen}, by making further use of the properties
of the R\'enyi entropies, it can be adapted to study questions that
Nielsen's theorem does not address, like the amount
of classical communication required to perform entanglement 
dilution~\cite{HaydenW02}.

The following lemma gives
\emph{dimension-independent} estimates of the R\'enyi entropy for
states that are close together. In contrast, the analogous result
for the von Neumann entropy, known as the Fannes inequality \cite{Fannes73},
has a logarithmic dependence on the rank of the
states.  

\begin{lemma}
\label{Elem:Renyi}
Let $F(\r,\s) > 1 - \e$. For all $1/2 \leq \alpha < 1$ the R\'enyi
entropies of order $\alpha$ of $\r$ are bounded below by
\begin{eqnarray*}
S_\alpha(\r) &\geq&  S_\b(\s) + \frac{2\a}{1-\a} \log(1-\e),
\end{eqnarray*}
where $\beta = \infty$ if $\alpha = 1/2$ and $\beta = \alpha /
(2\alpha - 1)$ otherwise.
\end{lemma}
\begin{proof}
It can be shown \cite{Fuchs95} that $F(\r,\s)$ is the
minimum over all measurements $\{M_i\}$ of the function
$\sum_i \sqrt{ \Tr(M_i \r) \Tr(M_i \s) }$.
Choosing $\{M_i\}$ to be projection in a basis that diagonalizes
$\r$ yields $F(\r,\s) \leq \sum_i (\l_i \om_i)^{1/2}$, where $\om_i
= \Tr(M_i \s)$ and $\{\l_i\}$ is the spectrum of $\r$.  When $\a >
1/2$, an application of H\"{o}lder's inequality, which states that
$\sum_i x_i y_i \leq (\sum_i x_i^p)^{1/p} (\sum_i y_i^q)^{1/q}$
when $x_i, y_i \geq 0$, $p > 1$ and $1/p + 1/q =  1$, 
with $p=2\a/(2\a-1)$ and $q=2\a$,  yields
\begin{eqnarray*}
\log(1-\e)
&\leq&
\frac{2\a-1}{2\a}\log \sum_i \om_i^{\a/(2\a-1)} +
\frac{1}{2\a} \log\sum_i\l_i^{\a}
\end{eqnarray*}
upon taking logarithms on both sides.  Because the $\om_i$
are outcome probabilities of a projective measurement, the vector
$(\om_i)$ is majorized by the spectrum of $\s$ \cite{AlbertiU82}.
The Schur concavity of the R\'enyi entropies
then implies that $S_\b(\om_i) \geq S_\b(\s)$, completing
the proof of the lemma for $\alpha \neq 1/2$.  The $\alpha = 1/2$
case follows by taking limits.
\end{proof}

Because the previous result makes no reference to the rank of the
states $\r$ and $\s$, it can be combined with lemma~\ref{Elem:exactRenyi}
to yield a family of bounds on the transformation complexity in the
presence of unlimited amounts of entanglement.  These bounds, in turn,
provide a convenient way to derive bounds on the quantum
communication complexity of distributed function evaluation.
\begin{theorem}
\label{Ethm:approxRenyi}
For any bipartite pure state $\ket{\ph_{AB}}$ and
$\ket{\ps_{AB}}$, and any $1/2\leq \a < 1$, the transformation
complexity obeys the inequality
\begin{eqnarray}
\label{Eeqn:noDimension}
\bar{Q}^{(*)}_\e(\ps_{AB}|\ph_{AB})
&\geq& S_\b(\ps_A) - S_\a(\ph_A) + \frac{2\a}{1-\a} \log(1-\e),
\end{eqnarray}
where $\beta = \infty$ if $\alpha = 1/2$ and $\beta = \alpha /
(2\alpha - 1)$ otherwise.
\end{theorem}
\begin{proof}
Suppose that there exists a maximally entangled state 
$\ket{\Ph^k_{AB}} := \smfrac{1}{\sqrt{k}}\sum_{j=1}^k\ket{j,j}$ 
and an $n$-qubit protocol for transforming $\ket{\ph_{AB}}\ox\ket{\Ph^k_{AB}}$
into the state $\ket{\r_{AB}}$ where 
$F(\ps_{AB}\ox\Ph^k_{AB},\r_{AB}) \geq 1-\e$.  By
the monotonicity of the fidelity, it then follows that also 
$F(\ps_A \ox \Ph^k_A,\r_A) \geq 1-\e$, and thus, according to 
lemma~\ref{Elem:Renyi}, $S_\a(\r_A) \geq S_\b(\ps_A)+\log k + \smfrac{2\a}{1-a}\log(1-\e)$.
Using $S_\a(\ph_A\ox\Ph^k_A) = S_\a(\ph_A) + \log k$, lemma~\ref{Elem:exactRenyi}
proves that the conversion from $\ph_{AB}\ox\Ph^k_{AB}$ to $\r_{AB}$ requires
at least $S_\b(\ps_A) - S_\a(\ph_A) + \frac{2\a}{1-\a} \log(1-\e)$ qubits
of communication, regardless of the dimension $k$.
\end{proof}
Note that the above proof relies critically on the fact that 
for the maximally entangled state $S_\b(\Ph^k_A) - S_\a(\Ph^k_A)=0$, 
which does not hold for general states.

It should be pointed out that similar lower bounds can be derived 
from other other techniques from quantum information theory, 
such as those of Ref.~\cite{VidalJN00}.
In practice, however, the ease with which the R\'enyi
entropies are calculated and manipulated can result in significant
simplification over approaches based on majorization.

\section{Quantum Lower Bounds on Communication Complexity}
In this section will use our results on the complexity of
 state transformations to obtain similar lower bounds on 
the communication complexity of functions.

\begin{definition}
For a function $f:X\times Y\rightarrow \{-1,+1\}$ 
the rectangle $R_f$ is defined by $R_{xy} := f(x,y)$
for all $x\in X$ and $y\in Y$.
The two \emph{marginal density matrices} of the 
function $f$ are $\rho_f^X := \smfrac{1}{|X||Y|}R_f R_f^T$ 
and $\rho_f^Y := \smfrac{1}{|X||Y|}R_f^TR_f$.
The \emph{eigenvalue spectrum} $\sigma(f)$ of $f$
is the spectrum of the marginal density matrix of $f$
(the spectra of $\rho_f^X$ and $\rho_f^Y$ are the same).
\end{definition}
It is easy to prove that the bipartite state
\begin{eqnarray}
\ket{\ps_{AB}} & = & \frac{1}{\sqrt{|X||Y|}}
\sum_{x\in X,y\in Y}{f(x,y)\ket{x,y}}
\end{eqnarray} 
will have marginal density matrices 
$\rho_f^X$ and $\rho_f^Y$, 
and hence $S_\b(\sigma(f)) = S_\b(\ps_A) = S_\b(\ps_B)$.

These definitions allow us to formulate the following theorem.
\begin{theorem}\label{thm:qcc}
Let $f:X\times Y\rightarrow \{-1,1\}$ be a distributed function.
The communication complexity 
$Q_{\e}^{(*)}(f)$ is lower bounded by
\begin{eqnarray}
Q_{\e}^{(*)}(f) & \geq & \smfrac{1}{2}S_\b(\sigma(f))+\smfrac{\b}{\b-1}\log(1-2\e),
\end{eqnarray}
for every $\beta > 1$.
This lower bound applies to the uniform distribution 
$\mu(x,y) := 1/{|X||Y|}$ over the input values.
\end{theorem}
\begin{proof}
We know by lemma~\ref{Elem:transition} that if we define 
$\ph$ and $\ps$ according to 
\begin{eqnarray} \label{eq:fstates}
\ket{\ph_{AB}} &=&\frac{1}{\sqrt{|X||Y|}} \sum_{x\in X,y\in Y} 
\ket{x_A} \ket{y_B},\\
\ket{\ps_{AB}} &=& \frac{1}{\sqrt{|X||Y|}} \sum_{x\in X,y\in Y} f(x,y) \ket{x_A} \ket{y_B},
\end{eqnarray}
we have $Q_\e^{(*)}(f) \geq \smfrac{1}{2}\Q^{(*)}_{2\e}(\ps_{AB}|\ph_{AB})$.
Because $\ph_{AB}$ is a disentangled state with $S_\a(\ph_A)=0$
and $S_\b(\ps_A) = S_\b(\sigma(f))$, we use theorem~\ref{Ethm:approxRenyi}
to complete the proof.
\end{proof}

\subsection{The Inner Product Function}

As an example, we will apply the lower bound of the previous
section to determine the quantum communication complexity of the
inner product function, which is defined for two $n$-bit 
strings $x,y\in\{0,1\}^n$ by
\begin{eqnarray}
\IP(x,y) & := & \sum_{i=1}^n x_i y_i \quad (\mbox{mod 2}).
\end{eqnarray}
In the exact case with prior entanglement, 
it has been shown that
$Q_0^*(\IP) = \left\lceil \smfrac{n}{2} \right\rceil$~\cite{ClevevNT99}.
The speedup of a factor of $2$ is achieved using superdense coding,
which is applicable to every function.  In the bounded-error case,
however, the known bounds are considerably weaker.  In the same
paper, the lower bound
$Q_\e^*(\IP) \geq \smfrac{1}{2} (1-2\e)^2 n - \smfrac{1}{2}$,
has a scaling factor in front of the linear term $n$ that depends 
on the error rate $\e$.  
This is in sharp contrast with the lower bound for the model
where no prior entanglement is allowed \cite{Ambainis98}:
$Q_\e(\IP) \geq \smfrac{n}{2} + \log(1-2\e)$.

Here we will generalize this inequality to the setting where 
$A$ and $B$ are allowed to use an arbitrary amount of EPR-pairs.
\begin{theorem}
\label{Ethm:QCCIP}
The EPR-enhanced communication complexity of the inner
product function is bounded by
\begin{equation}
\smfrac{n}{2} + \log(1-2\e)
\quad \leq\quad Q_\e^{(*)}(\IP) \quad \leq \quad 
\max \left( 0, \left\lceil \smfrac{n}{2} + \smfrac{1}{2} \log(1-2\e) \right\rceil \right).
\end{equation}
\end{theorem}
This result, we should emphasize, only holds if the entangled
state shared by Alice and Bob is restricted to be of the form
$\ket{\Ph^k} = \smfrac{1}{\sqrt{k}} \sum_{j=1}^k \ket{j,j}$.
In independent work, Nayak and Salzman, using a different technique 
than the one we
have employed here, have recently shown that the lower bound actually
holds for $Q_\e^*(\IP)$ as well \cite{NayakS}.  

To prove the lower bound we use our earlier theorem~\ref{thm:qcc}
in combination with the knowlegde that the rectangle $R_{\mathrm{IP}}$
is an orthogonal matrix such that $\rho^X_{\mathrm{IP}} = \smfrac{1}{2^n}I$.
As a result, $S_\b(\sigma(\IP))=n$ for all $\b$ and the lower bound
follows by taking the limit $\b \rightarrow \infty$.

For the upper bound consider $q$ qubits of communication used in
a superdense coding scheme such that Alice can send Bob $2q$ bits
of information about her $n$-bit input string. Bob can then guess
the value of $x$ on Alice's remaining $n-2q$ bits, expecting to be
correct with probability $2^{2q-n}$. On the occasions that he has
guessed correctly, he can evaluate $\IP$ without error, otherwise,
he can expect to be correct half the time. Therefore, the
probability of success for the resulting protocol will be
$1 - \e \geq 2^{2q-n} + \frac{1}{2}(1-2^{2q-n})$.
Taking logarithms then reveals that for this protocol:
$q \leq \frac{n}{2} + \frac{1}{2}\log(1-2\e)$.

Observe also that any protocol that uses prior entanglement and 
classical communication (with complexity measure $C^*_\e(\IP)$) 
can be simulated at half the cost
using superdense coding and that the optimal quantum protocol
constructed here simply involves one-way communication from Alice
to Bob.  It follows that 
$C^*_\e(\IP) \geq 2 Q_\e^{(*)}(\IP) = n+2\log(1-2\e)$.


\subsection{Lower Bound for a Promise Function}
Consider the distributed function $g:\FF_q\times\FF_q \rightarrow \{-1,+1\}$
that is defined for all $x\neq y$ as the \emph{quadratic character} 
of the difference between $x$ and $y \in\FF_q$: 
\begin{eqnarray}\label{eq:quadchar}
g(x,y) & := & \left\{
\begin{array}{rl}
+1 & \mbox{ if $x-y$ is a square in $\FF_q$,}\\
-1 & \mbox{ if $x-y$ is not a square in $\FF_q$.}
\end{array}
\right.
\end{eqnarray}
(See \cite{IrelandR90} for the various properties of quadratic characters.)
Here we will give an almost tight lower bound 
on the communication complexity $Q^{(*)}_\e$ of this function 
using the R\'enyi entropic methods of the previous sections. 

First we should note that $g$ is not defined on all input pairs $(x,y)$,
which means that $g$ is an example of a `promise function'.
Because of this, we cannot directly apply the lower bound method of 
theorem~\ref{thm:qcc} as this assumes the uniform prior distribution
over the input states.  Instead, we will have to prove our lower bound
for the correlated distribution $\mu(x,y) = \smfrac{1}{q(q-1)}$ for
$x\neq y$, and $\mu(x,x)=0$.  Fortunately this is possible with the help
of theorem~\ref{Ethm:approxRenyi}, which can deal with initial entangled 
states $\ph_{AB}$. 
\begin{theorem}
Let $g:\FF_q\times\FF_q\rightarrow \{-1,+1\}$ 
be the function defined in Equation~\ref{eq:quadchar}.
The EPR-enhanced communication complexity of this quadratic character
function is bounded 
\begin{equation}
\log(q-1) - 2 + 2\log(1-\e) \quad \leq \quad Q^{(*)}_\e(g) \quad \leq \quad 
\lceil \log(q)\rceil.  
\end{equation} 
Specifically, the lower bound applies to the distribution: 
$\mu(x,y) = \smfrac{1}{q(q-1)}$ for $x\neq y$ and 
$\mu(x,x)=0$ over the $q(q-1)$ input pairs $(x,y)$. 
\end{theorem}

\begin{proof}
The upper bound is trivial: the set $\FF_q$ has $q$ elements, hence 
$\lceil \log q \rceil$ bits are sufficient for Bob to transmit 
the value of $y$ to Alice.

For the lower bound, define the initial and final states $\ph$ and $\ps$
\begin{eqnarray} \label{eq:gstates}
\ket{\ph_{AB}} &=&\frac{1}{\sqrt{q(q-1)}} \sum_{x\neq y\in \FF_q} 
\ket{x_A} \ket{y_B},\\
\ket{\ps_{AB}} &=& \frac{1}{\sqrt{q(q-1)}} \sum_{x\neq y\in \FF_q} g(x,y) \ket{x_A} \ket{y_B}.
\end{eqnarray}
For the quadratic character the following `shift property' holds (see \cite{IrelandR90}):
\begin{eqnarray}
\sum_{x\in\FF_q}{g(x+r)g(x+s)} & =& 
\left\{
\begin{array}{rl}
-1 & \mbox{ if $s\neq r$,}\\
q-1 & \mbox{ if $s=r$.}
\end{array}
\right.
\end{eqnarray}
It is thus not hard to see that the marginal density matrices of 
the two states are $\ph_A = \smfrac{1}{q(q-1)}[I+(q-2)J]$ and 
$\ps_A = \smfrac{1}{q(q-1)}[qI-J]$, where $I$ is the $q$-dimensional
identity matrix, and $J$ is the `all ones' matrix of the same 
dimension.  The respective spectra are thus
$\sigma(\ph_A) = (1-\smfrac{1}{q},\smfrac{1}{q(q-1)},\dots,\smfrac{1}{q(q-1)})$
and $\sigma(\ps_A) = (\smfrac{1}{q-1},\dots,\smfrac{1}{q-1},0)$.
With this knowledge we can employ the lower bound of Theorem~\ref{Ethm:approxRenyi}.
For $(\alpha,\beta) = (\smfrac{1}{2},\infty)$ 
the entropies of the two spectra are $S_\infty(\ps_A) = \log(q-1)$
and $S_{1/2}(\ph_A) = \log(4-\smfrac{4}{p}) < 2$, and hence indeed  
$\bar{Q}^{(*)}_\e(\ps_{AB}|\ph_{AB}) \geq \log(q-1) - 2 + 2\log(1-\e)$.
\end{proof}

\section*{Acknowledgments}
WvD is supported
by an HP/MSRI postdoctoral fellowship, the Defense 
Advanced Research Projects Agency (DARPA) and the Air Force Laboratory, Air 
Force Materiel Command, USAF, under agreement number F30602-01-2-0524.
PH is supported by a Sherman Fairchild fellowship and US
National Science Foundation grant no.~EIA-0086038.

\end{document}